\documentclass{aastex}
\usepackage{emulateapj5}

\submitted{Submitted 20 March 2002; Accepted 11 September 2002 -- To appear in the December, 2002 AJ}
\newcommand{\etal}{{\it et\thinspace al.}\ }
\lefthead{Stevenson \etal}
\righthead{KISS AGNs}

\begin{document}

%\slugcomment{Revised version: August 14, 2002}

\title{KISS AGNs in the Soft X-ray Band: 
\\Correlation with the ROSAT All-Sky Survey}

\author{Samantha L. Stevenson, John J. Salzer}
\affil{Astronomy Department, Wesleyan University, Middletown, CT 06459; sam@astro.wesleyan.edu, slaz@astro.wesleyan.edu}
\author{Vicki L. Sarajedini}
\affil{Astronomy Department, University of Florida, Gainesville, FL 32611; vicki@astro.ufl.edu}
\and
\author{Edward C. Moran}
\affil{Astronomy Department, Wesleyan University, Middletown, CT 06459; ecm@astro.wesleyan.edu}

\begin{abstract}
We present a study of the X-ray properties of a volume-limited sample of optically selected emission-line galaxies. The sample is derived from a correlation between the KPNO International Spectroscopic Survey (KISS), an $H\alpha$-selected objective-prism survey of AGNs and starbursting galaxies, and the ROSAT All-Sky Survey (RASS). After elimination of all spurious matches, we identify 18 ROSAT-detected X-ray sources within the KISS sample in the $0.1-2.4$ keV band. Due to soft X-ray selection effects, the majority of the ROSAT sources are Seyfert 1 galaxies.  The majority (54\%) of the ROSAT-KISS Seyferts are classified as narrow-line Seyfert 1 objects, a relatively high percentage compared to previous objective-prism-selected Seyfert galaxy samples.  We estimate the X-ray luminosities of the ROSAT-detected KISS objects and derive volume emissivities of $6.63 \times 10^{38} \mathrm{ergs}$ $ \mathrm{s}^{-1} \mathrm{Mpc}^{-3}$ and $1.45 \times 10^{38} \mathrm{ergs}$ $ \mathrm{s}^{-1} \mathrm{Mpc}^{-3}$ for the $30^{\circ}$ and $43^{\circ}$ strips, respectively. For those KISS AGNs not detected by RASS, we use the median $L_X/L_{H\alpha}$ ratio derived from a previous study to estimate $L_X$. The total $0.5-2$ keV volume emissivity we predict for the overall KISS AGN sample is sufficient to account for $22.1 \pm 8.9\%$ of the soft X-ray background (XRB), averaged over both survey strips. The KISS AGN sample is made up predominantly of intermediate-luminosity Seyfert 2's and LINERs, which tend to be weak soft X-ray sources. They may, however, represent a much more significant contribution to the hard XRB.
\end{abstract}


\section{Introduction}
Active galactic nuclei (AGNs) are among the most energetic objects in the universe, with energy output spanning the entire electromagnetic spectrum. It is currently believed that the majority of soft X-ray emission by AGNs arises from superheating of the accretion disk surrounding the central supermassive black hole. However, the X-ray emission actually observed from AGNs depends strongly on the degree to which the ``central engine'' is obscured by surrounding material (e.g., the dusty torus model). The observed spectrum (i.e., hardness) of the source also depends on the amount of obscuration present. This paper seeks to study the soft X-ray properties of cosmologically local ($z \leq 0.095$) AGNs. Towards this end, we have constructed an optically selected sample of such AGNs detected as part of the ROSAT All-Sky Survey (RASS).

 In order to better understand the X-ray emission properties of local AGNs, we have compiled a relatively unbiased sample of these objects for study. The KPNO International Spectroscopic Survey (KISS; Salzer et al. 2000, 2001) provides a unique sample of galaxies ideally suited for statistical analyses. The KISS sample includes all emission-line galaxies (ELGs) with detectable H$\alpha$ emission (EW $\geq 20-30$\AA). Since it is a line-selected sample with an imposed maximum wavelength, there is an implicit redshift limit associated with the survey. This limit, along with the great sensitivity of the survey, means that KISS is effectively volume-limited for the more luminous portion of the ELG sample, which includes the majority of the AGNs.

Because of the unique way in which KISS galaxies are selected, our ROSAT-KISS comparison differs in a fundamental way from previous X-ray/optical correlation studies. Most of these are X-ray selected (e.g., Ueda et al. 1999, Griffiths et al. 1996) and focus on the optical properties of objects detected in the X-ray. Here, we are studying the X-ray properties of an optically selected sample. This approach yields a sample of galaxies significantly different from those derived by previous authors (see Section \ref{Properties}).

Studies by \citet{h01} and \citet{ho01} also use optical catalogs as the primary basis for selection. Their galaxy samples are drawn from the Palomar survey of nearby galaxies, a magnitude-limited survey which has full coverage of the northern sky. The sample constructed from this survey is volume-limited and complete to a comoving distance of $\approx$ 18 Mpc. Although the sky coverage of KISS is only 128 $\mathrm{deg}^2$ in total, the sample volume is quite deep: the comoving distance limit of the survey is $\sim 343$ Mpc. The current study complements those of Halderson et al. and Ho et al. by exploring the X-ray properties of a much more distant (but still cosmologically ``local'') sample. On average, the ROSAT-KISS sample is $\approx$ 100 times more luminous than the Halderson et al. sample in the soft X-ray band. 

Emission-line galaxies similar to those discovered by KISS are important
components of the cosmic X-ray background (XRB).  The XRB was discovered
40 years ago \citep{g62} in the 2--6 keV range, and has since been
demonstrated to extend to both higher and lower energies (see the review by 
Fabian \& Barcons 1992).  In deep images obtained recently with the {\it Chandra
X-ray Observatory}, $\sim$~80--90\% of the background in the soft (0.5--2 keV)
and hard (2--10 keV) X-ray bands has been resolved into discrete sources
(e.g., Rosati et al.\ 2002).  Models (Comastri et al.\ 1995; Gilli et al.\
2001) and observations of both nearby (Moran et al.\ 2001) and distant
(Tozzi et al.\ 2001) X-ray sources have indicated that the majority of the
hard XRB is produced by type~2 AGNs whose soft X-ray fluxes are absorbed
by dense circumnuclear material.  In contrast, at least 50\% of the XRB
at lower energies is produced by high-luminosity, unobscured (type~1) AGNs
and quasars (Hasinger et al.\ 1998; Schmidt et al.\ 1998).  A significant
fraction of the remaining soft XRB appears to be associated with optically
bright galaxies of relatively low X-ray luminosity (Hornschemeier et
al.\ 2001; Tozzi et al.\ 2001).  It is unclear whether these objects are
low-luminosity AGNs, non-active galaxies, or a combination of the two.  A
major aim of this study, therefore, is to constrain the soft XRB contribution
of moderately luminous AGNs.

We describe the KISS and ROSAT catalogs and the results of the X-ray-optical matching in Section 2. The properties of the galaxies in the correlation sample are discussed in detail in Section 3. We present XRB calculations in Section 4; we take into account contributions from both ROSAT-KISS objects and those whose 0.1-2.4 keV emission lies below the ROSAT detection threshold. We close with a discussion of the implications of our analysis in the context of current XRB research.

\section{Optical/X-ray Source Populations}

\subsection{KPNO International Spectroscopic Survey - KISS}
\label{KISS_sum}

KISS is a line-selected, objective-prism survey for extragalactic emission-line objects \citep{salzer00}. KISS is the first purely digital spectroscopic survey for emission-line galaxies (ELGs). It therefore represents a significant advance beyond slitless spectroscopic surveys carried out using photographic plates. Examples of such photographic surveys include the Markarian \citep{mark67, mark81}, Michigan (UM; \citet{mac77, mac81}), Case \citep{pesch83, st92} and UCM \citep{zam96} surveys. Its digital nature allows KISS to reach much fainter flux limits than previous studies, and therefore to reach a substantially greater galaxian surface density (18.1 $\mathrm{deg}^{-2}$ over the $30^{\circ}$ strip; see below).

KISS is being carried out over several ``strips'', each $\approx 1^{\circ}$ wide in declination \citep{salzer00}. The first survey area (KR1; hereafter the $30^{\circ}$ strip) covers the region RA = 12h 15m - 17h 0m, $\delta$ = $29^{\circ}$ - $30^{\circ}$, coinciding with part of the CfA Century Redshift Survey \citep{geller97, weg01}. The second (KR2; hereafter the $43^{\circ}$ strip) runs through the Bootes void: RA = 11h 55m - 16h 15m, $\delta$ = $42.7^{\circ}$ - $44.3^{\circ}$. The total sky coverage of these two survey regions is 128.1 $\mathrm{deg}^2$. In the $30^{\circ}$ strip, 1128 ELGs have been cataloged \citep{salzer01}, while a total of 1029 have been identified in the $43^{\circ}$ strip \citep{gron02a}. To date, high-resolution follow-up spectra have been obtained for 64\% of the objects in the $30^{\circ}$ strip and 14\% of those in the $43^{\circ}$ strip, allowing precise determination of redshift and ELG activity type (i.e., starburst vs. AGN). 

Survey data were obtained in two distinct spectral regions. The blue survey (Salzer et al. 2002) selects ELGs by the presence of the [O III] $\lambda$5007 emission line; the red survey (see Salzer et al. 2001) selects \emph{via} the H$\alpha$ line. This paper focuses on the red spectral data: only the H$\alpha$-selected portion of the KISS survey data is considered. The spectral range of the red survey data extends from 6400 to 7200 \AA, beginning slightly below the rest wavelength of the H$\alpha$ line and cutting off before the OH molecular band at about 7240 \AA. This implies that the H$\alpha$ line is detectable to a redshift of $z=0.095$. In a minority of cases a high-redshift emission-line object will be detected by KISS when another strong spectral line (e.g., [O III], H$\beta$) redshifts into the KISS wavelength range. Several QSOs and Seyfert galaxies have been detected by KISS in this manner. 

Because of its well-defined redshift cutoff, KISS is essentially volume-limited for the more luminous portion of the sample. Some incompleteness arises for galaxies with faint H$\alpha$ line strengths. However, the majority of the KISS AGNs are included in the volume-limited portion of the sample. For a detailed discussion of KISS completeness limits, see \citet{gron02c}. Due to its depth and emission line selection method, KISS is ideal for performing relatively unbiased statistical studies of the local population of star-forming galaxies and AGN.

\subsection{ROSAT All-Sky Survey}
 \label{ROSAT_sum}

The RASS resulted in a sample of soft (0.1-2.4 keV) X-ray sources which is a factor of 20 deeper than previous X-ray surveys such as the Einstein Medium Sensitivity Survey \citep{gioia90}, and which has nearly full sky coverage. At a brightness limit of 0.1 cts/s, the survey covers 92\% of the sky \citep{v99}. The large effective area and short focal length of ROSAT, combined with the low background noise of the PSPC, resulted in an instrument that was designed to study X-ray point sources, but which was nearly ideal for studying diffuse X-ray emission as well. As a result, the survey yielded the most accurate maps to date of the diffuse XRB in the 0.1-2.4 keV energy band \citep{sn95}. These maps have an angular resolution of approximately $2^{\circ}$. The resolution of these maps was later improved to 12' \citep{sn97}. Deep surveys later conducted by ROSAT \citep{h98} revealed that 70 to 80\% of this diffuse background could in fact be resolved into discrete sources to flux levels on the order of $1\times 10^{-15}$ $\mathrm{erg}$ $\mathrm{s}^{-1}$ $\mathrm{cm}^{-2}$. 

The RASS was performed using the Position Sensitive Proportional Counter (PSPC), an instrument with a positional accuracy on the order of 10 arcsec (correlation with the TYCHO catalog of bright stars shows that 68\% of Bright Source Catalog objects are found within 13 arcsec of the optical position; Voges et al. 1999). Catalogs derived from RASS provide detailed information about each source, including position, positional error, source count rate and error, and exposure time. In constructing the ROSAT-KISS database, we have combined all relevant X-ray source information with the optical data contained in the KISS tables \citep{salzer01, gron02a}. The X-ray data used in this paper is drawn from the Bright and Faint Source Catalogs, which are derived directly from processing of the All-Sky Survey data \citep{v99}. 

 The Bright Source Catalog (BSC) consists of the brightest objects in the RASS-II database (resulting from the second RASS processing; see Voges et al. 1999). The limiting PSPC count rate is $0.05$ cts/s with at least 15 source counts required for inclusion, or a detection likelihood of at least 15. Detection likelihood is defined as $-\ln(P)$, where P is the probability that the observed photon distribution is due to random background fluctuations. These two criteria result in a source catalog of 18,811 objects.

The second and larger catalog constructed from All-Sky Survey data is the Faint Source Catalog (FSC). The FSC is an extension of the BSC to fainter limiting count rates; it represents the full measure of the non-spurious sources detected in the RASS-II processing, containing 145,060 objects. The criterion for inclusion in the FSC is at least 7 source counts, with a detection likelihood of 7 or greater. The median count rate for FSC sources in the two KISS areas is approximately $0.02$ cts/s. (See Section \ref{Predictions} for discussion of the FSC flux threshold.) 

Previous RASS correlations have been performed with catalogs of sources at various wavelengths. These include correlations with the ROSAT catalog of pointed observations (WGACAT; White et al. 1994), the SIMBAD and NED public databases, the FIRST radio survey \citep{brink00}, and the IRAS catalogs \citep{boll98}.

\subsection{The Matching Process}
A positional-coincidence test was used to conduct the correlation between the KISS database and the BSC/FSC, based on correspondences between the coordinates of objects in X-ray and optical wavelengths. A uniform search radius of 60 arcsec was adopted: all ROSAT sources within 60 arcsec of a KISS ELG were considered to be possible matches. The median formal uncertainty in ROSAT source position recorded in the BSC/FSC for ROSAT-detected KISS objects is $\approx$ 10 arcsec, with errors as large as 25 arcsec present in the final dataset. 

There were a total of 22 matches in the first-pass correlation table, which includes all matches, legitimate and spurious, for ROSAT BSC and FSC sources appearing within 60 arcsec of a KISS object. The median difference between  X-ray and optical source positions was 10.9 arcseconds, with a standard deviation of 12.0; any source with a positional offset greater than 30 arcseconds was therefore flagged as a possible spurious match. Each source pair was verified visually using the SkyView\footnote{http://skyview.gsfc.nasa.gov/} database, to rule out chance correspondences. X-ray contour images drawn from the archive and overlaid on the online Digitized Sky Survey (DSS) images were used to confirm positional matches\footnote{In performing visual comparisons, we have made use of the ROSAT Data Archive of the Max-Planck-Institut f$\mathrm{\ddot{u}}$r extraterrestrische Physik (MPE) at Garching, Germany.}.  For a number of sources the KISS object is too faint to be visible on the DSS images.  In these cases the location of the X-ray contours relative to other objects in the DSS field was compared with the finder charts available for all KISS ELGs (which are substantially deeper than the DSS images) in order to verify the match.  Three KISS emission-line galaxies with positional offsets of greater than 30 arcsec were discarded after visual examination (KISSR 624, 662, and 2128, with 57.2, 33.7 and 48.8 arcsec positional offsets, respectively).  Another matched galaxy, KISSR 229, was found not to be an ELG after acquisition of a follow-up spectrum. Furthermore, the positional offset between the ROSAT source and KISSR 229 is 24.7 arcsec. Thus, KISSR 229 is not likely to be the true optical counterpart for this X-ray source, since the high density of non-ELGs in these fields greatly increases the odds of chance correspondence with an X-ray source position.

The remaining 18 objects were judged to be likely matches after visual examination. The median X-ray/optical positional offset for these galaxies is 9.6 arcsec. This is roughly consistent with the results of the correlation by Brinkman et al. (2000) of the ROSAT and FIRST (20cm) catalogs. This study finds that only 2\% of all ROSAT-FIRST matches are chance coincidences, to an angular separation of 30 arcsec. 

The 18 objects described above comprise the ROSAT-KISS sample. Their distribution between the two KISS areas and the BSC/FSC are summarized in Table \ref{table1}, while their optical and X-ray characteristics can be found in Table \ref{table2}.

%\placetable{table1}

Comparisons were also made between the KISS catalogs and the ROSAT public archive of pointed PSPC observations (WGACAT). The correlation was conducted remotely via the HEASARC server\footnote{http://heasarc.gsfc.nasa.gov}, again using a search radius of 60 arcsec. Only two RASS-detected sources were found to have corresponding pointed observations: KISSR 281, a Seyfert 1 galaxy, and the nearby LINER KISSR 29 (NGC 4278; for details, see Section \ref{Properties}). Observations of the latter object can also be found in the public archive of ROSAT High Resolution Imager (HRI) images; further search of this latter catalog revealed that no other KISS galaxies had been detected. For the sake of uniformity, we use only the RASS PSPC fluxes for these two sources in our analysis.

%\placefigure{figure1}
\begin{figure*}[htp]
\vskip -0.5in
\epsfxsize=7.0in
\hskip -0.3in
\epsffile{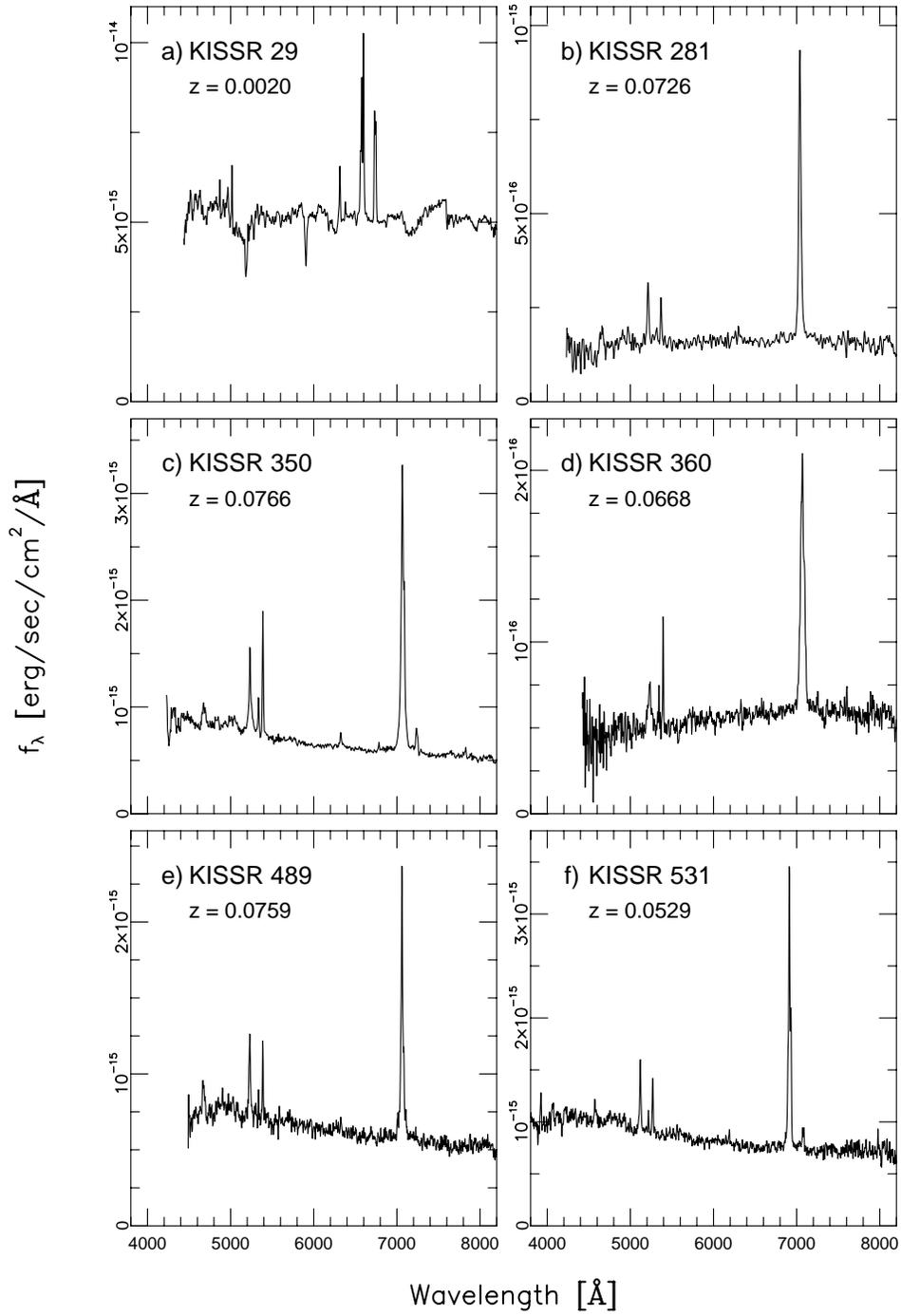}
\figcaption[/d4/kiss/specplots/sam_spec1.eps]{Follow-up spectra for the ROSAT-KISS objects.\label{figure1}}
\end{figure*}

\begin{figure*}[htp]
\vskip -0.5in
\epsfxsize=7.0in
\hskip -0.3in
\epsffile{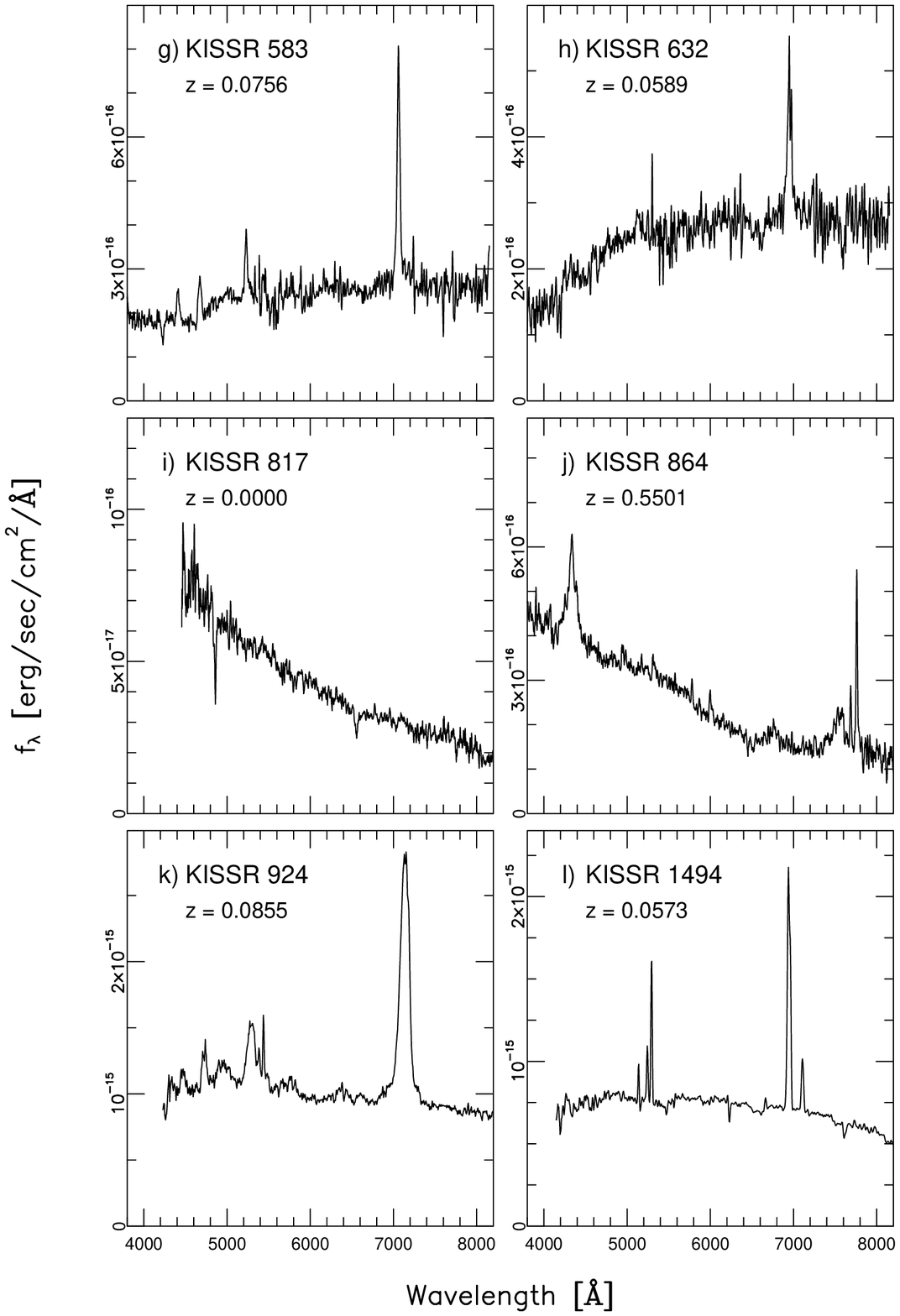}
\end{figure*}

\begin{figure*}[htp]
\vskip -0.5in
\epsfxsize=7.0in
\hskip -0.3in
\epsffile{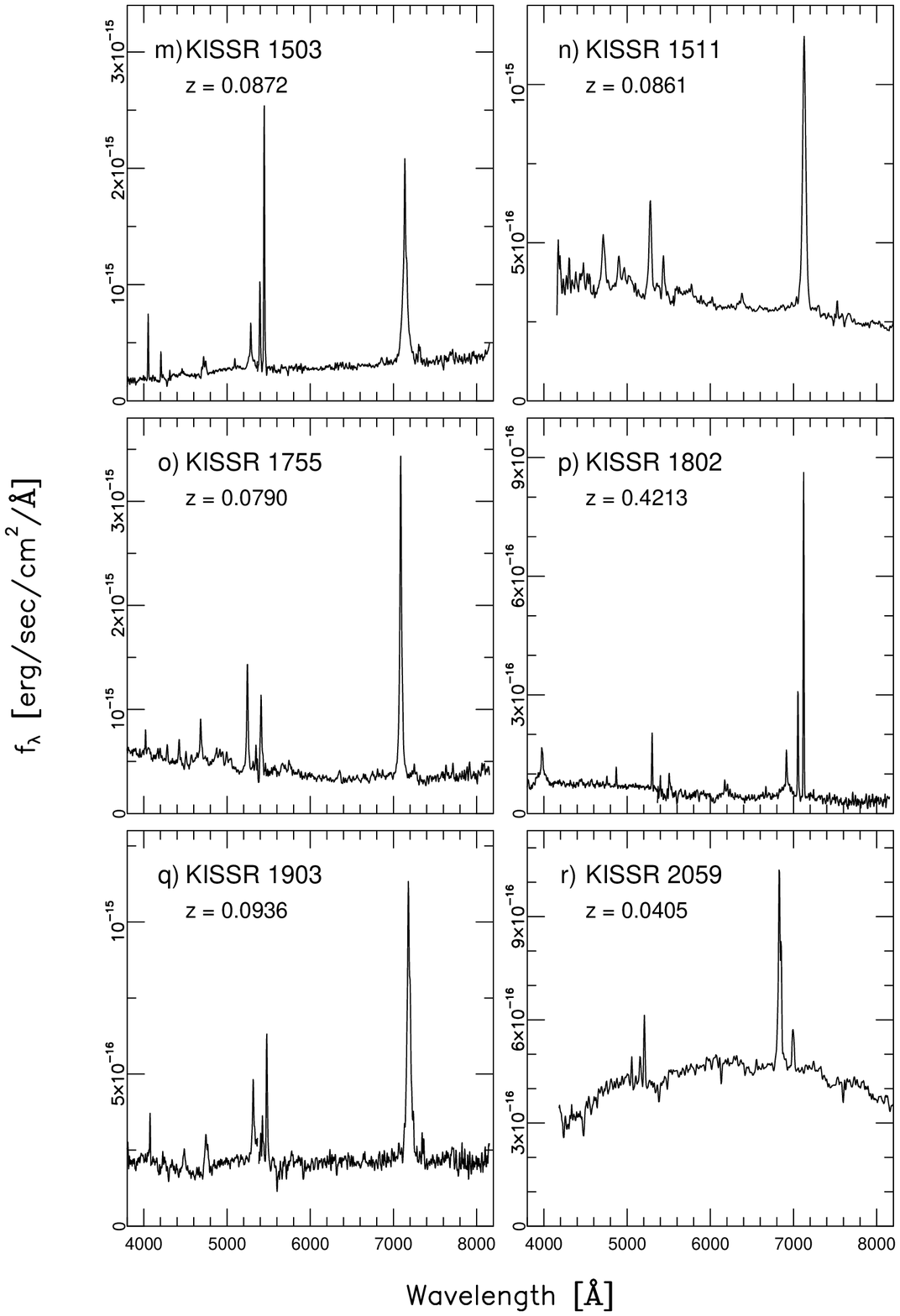}
\end{figure*}

%\placefigure{figure2}
\begin{figure*}[htp]
\vskip 0.2in
\epsfxsize=4.4in
\hskip 1.3in
\epsffile{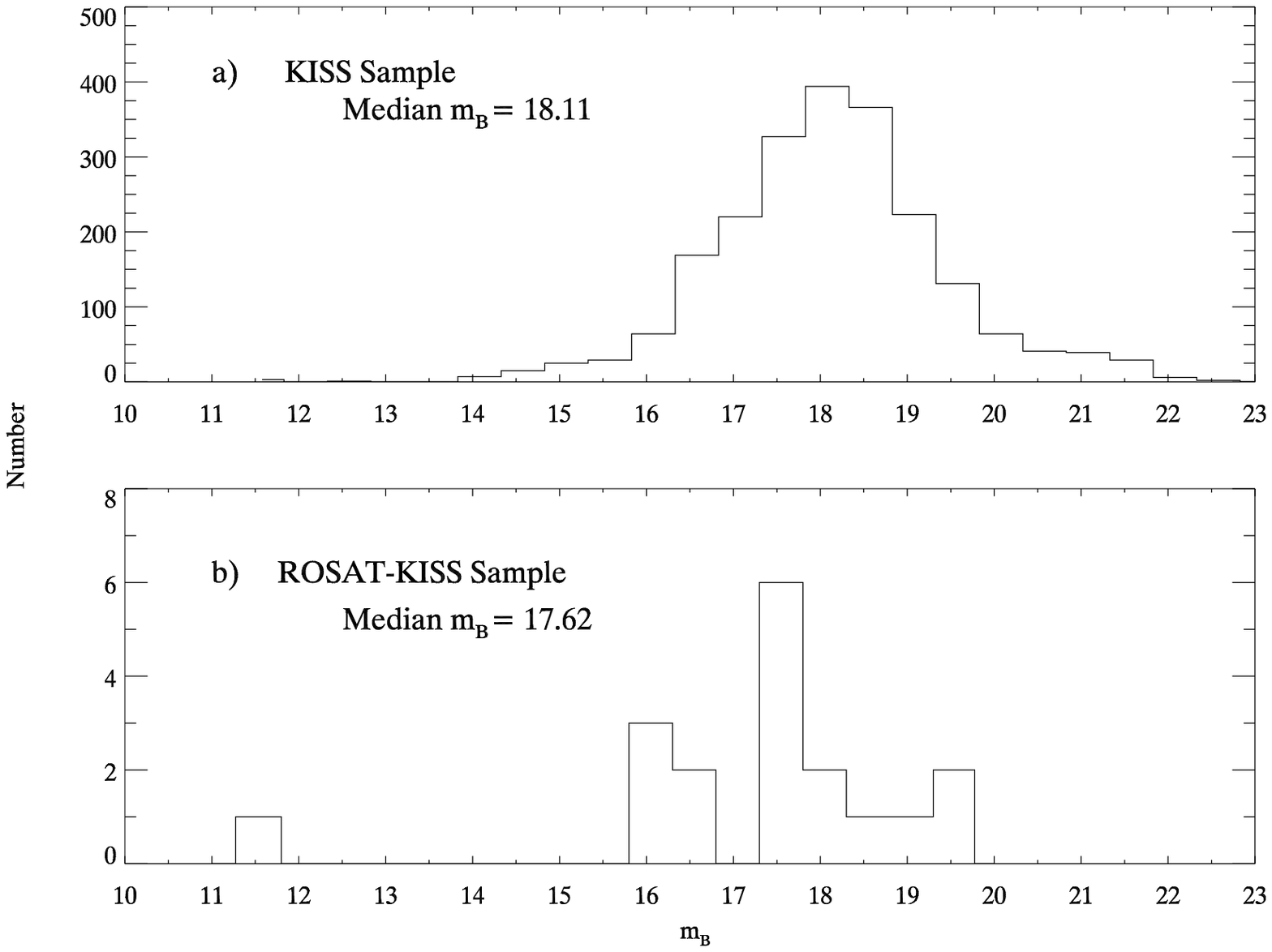}
\figcaption[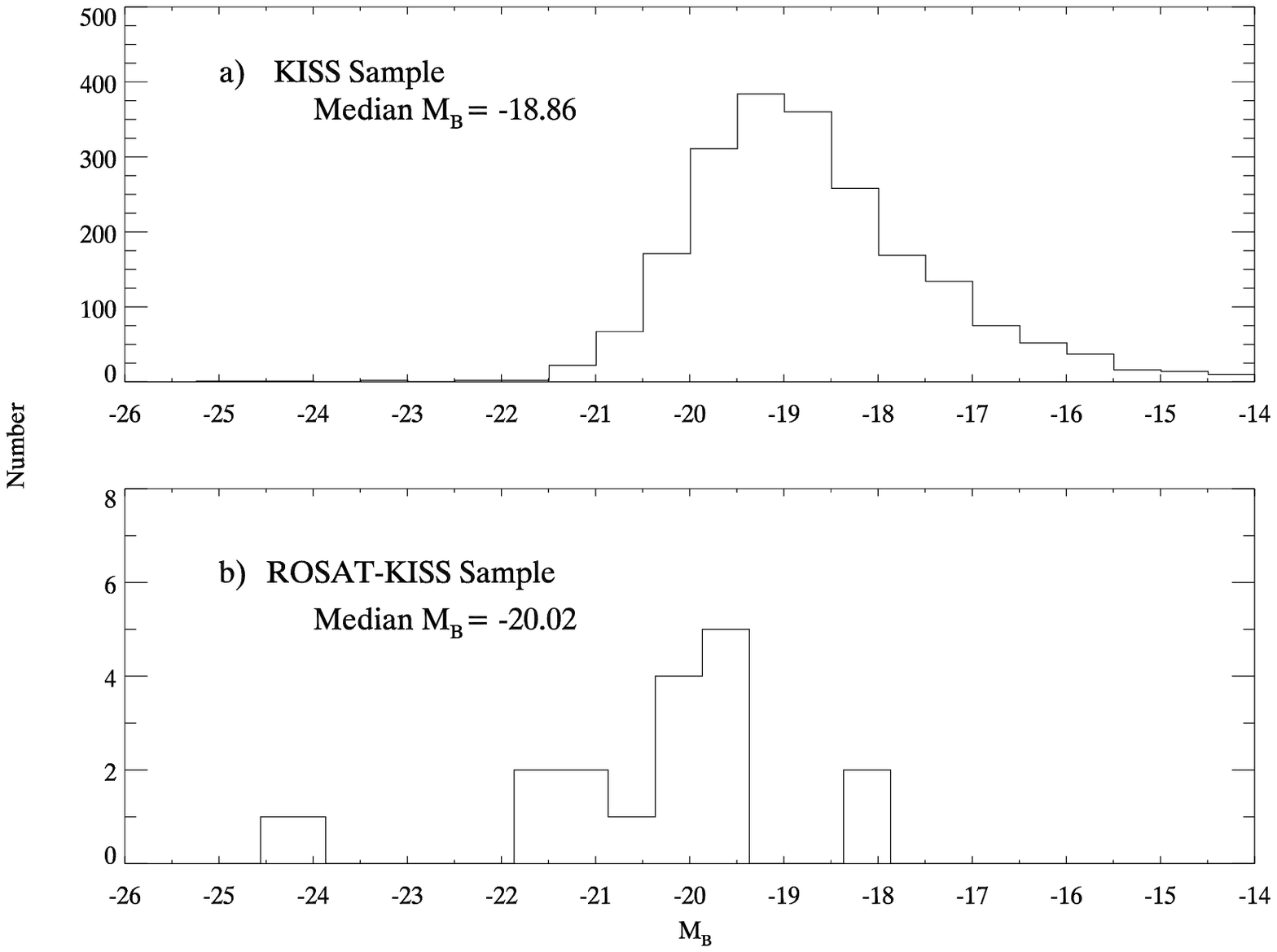]{B-band apparent magnitudes for a) the general KISS ELG sample and b) the ROSAT-KISS sample. The ROSAT-KISS objects are on average slightly brighter than the general KISS population. \label{figure2}}
\end{figure*}

\section{Properties of the Sample}
\subsection{Description of ROSAT-KISS objects}
\label{Properties}
The ROSAT-KISS sample consists of 18 X-ray detected sources. The basic properties of these objects are listed in Table \ref{table2}. Column 1 gives the KISSR catalog number of each source, while Column 2 lists the corresponding designation in the appropriate RASS catalog. Columns 3 and 4 list the RA and Dec coordinates of each source (equinox J2000), taken from the KISS catalogs \citep{salzer01, gron02a}. The symbol $\Delta$ in Column 5 of the table represents the offset in arcseconds between the optical and X-ray positions of the object. In Column 6, the ELG type is given for each object. The absolute and apparent B magnitudes of all ROSAT-KISS sources are listed in Columns 7 and 8. The B - V color of all objects is recorded in Column 9, while their redshifts can be found in Column 10. All photometric and spectroscopic data come from the KISS database. The PSPC count rate for each source is given in Column 11, while the 0.1-2.4 keV luminosities derived from this count rate (using appropriate spectral parameters; see Section 4.1) make up Column 12.

%\placetable{table2}

Follow-up optical spectra have been obtained for approximately 900 galaxies in the overall KISS sample, during the period from 1998-2001. Results of spectral observations are given in \citet{melb02}, \citet{weg02}, and \citet{salzer03}. The spectroscopic properties of the KISS ELG sample as a whole can be found in \citet{gron02b}.  Follow-up optical spectra are available for all 18 ROSAT-KISS objects, making activity type classifications (AGN/starburst) and accurate redshift determination possible in all cases. The spectra for all objects in the ROSAT-KISS sample are shown in Figure \ref{figure1}. Knowledge of the spectroscopic properties of the sample is extremely useful for establishing the nature of each source. Accurate redshift data is also essential in XRB calculations requiring knowledge of the emissivity within a certain volume (Section \ref{XRB}), where it becomes necessary to exclude luminosity contributions from sources outside the limiting H$\alpha$ redshift of KISS. 

The spectral characteristics of the X-ray selected sample vary widely, from stellar absorption lines to broad-line QSO emission. Thirteen of the 18 objects in the ROSAT-KISS sample have been spectroscopically identified as Seyfert 1 galaxies.  This high percentage of Seyfert 1s is most likely due to the fact that the ROSAT PSPC detection threshold corresponds to a relatively high flux level, meaning that only the brightest soft X-ray sources will be reliably detected. The low degree of obscuration associated with broad-line objects implies that they will be more readily detected in the soft X-ray band than will narrow-line AGNs. Four of the remaining five objects are other types of AGN, including one Seyfert 2 galaxy, one QSO, and two LINERs. The remaining source, KISSR 817, is identified as a stellar X-ray source (see Section \ref{indiv}). We note in passing that although KISS contains a sample of $\approx$ 2000 starbursting galaxies, none of these objects were detected by ROSAT.

\begin{figure*}[htp]
\vskip 0.2in
\epsfxsize=4.4in
\hskip 1.3in
\epsffile{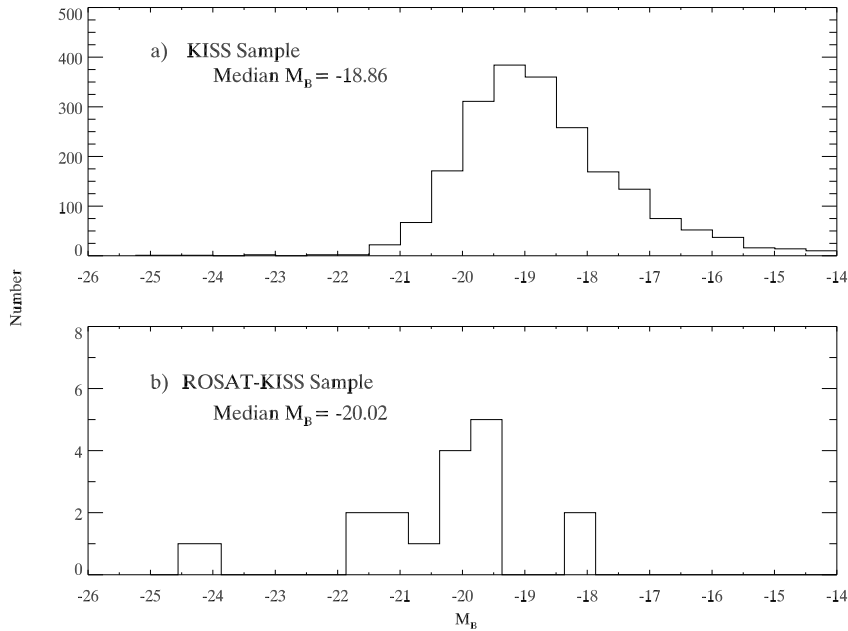}
\figcaption[/d7/kiss/rosat/paper/figures/absmaghist.eps]{B-band absolute magnitudes for a) the general KISS ELG sample and b) the ROSAT-KISS sample. The median absolute magnitude for the ROSAT-KISS objects is brighter than that of the general KISS sample by more than a full magnitude.\label{figure3}}
\end{figure*}

Figures \ref{figure2} and \ref{figure3} show the distribution of apparent and absolute magnitudes for both the general KISS sample and the ROSAT-KISS sample. The median in each case is brighter for the ROSAT-detected objects than for the general KISS objects. For the KISS sample as a whole, the median absolute magnitude $M_B = -18.86$. This is more than a full magnitude less luminous than the median for the ROSAT-KISS sample ($M_B = -20.02$). The same is true for apparent magnitude, where the median $m_B = 18.11$ for the general KISS population, but is $m_B = 17.62$ for the ROSAT-KISS objects. 

The general KISS ELG sample is heavily dominated by starbursting galaxies. For example, $88.4\%$ of the 663 KISS ELGs with follow-up spectra from the $30^{\circ}$ strip are starbursting galaxies. Only 5.3\% of these galaxies are Seyferts, and 5.7\% are LINERs. Of the Seyfert galaxies, 34.3\% exhibit broad lines (Seyfert 1s). There are only 19 Seyfert 1 galaxies identified to date in the whole of the sample. A large fraction (68\%) of the Seyfert 1s detected by KISS belong to the ROSAT-detected sample. These galaxies are among the most optically luminous objects in the KISS database (median $M_B = -20.15$). 

An interesting aspect of the ROSAT-KISS Seyfert 1 subsample
is the high proportion of ``narrow-line'' Seyfert 1 (NLS1) galaxies
present.  Inspection of Figure 1 reveals that many of the Seyfert 1's 
exhibit small FWHM values for both H$\beta$ and H$\alpha$, as well as 
substantial Fe II emission on either side of the H$\beta$ and [O III]
lines.  Examples include KISSR 281, 350, 489, 531, 1511, 1755, and 1903.
These galaxies all have H$\beta$ FWHMs of less than 1500 km/s and
[O~III]/H$\beta$ flux ratios of less than $\sim 3$, consistent
with the emission-line criteria for NLS1's described by Osterbrock
\& Pogge (1985) and Goodrich (1989).  NLS1's have generally steeper
soft X-ray spectra than classical Seyfert 1's (Boller, Brandt, \&
Fink 1996); as a result, previous soft X-ray selected samples of AGNs
have included high fractions of these objects (e.g., Stephens 1989;
Bade et al.\ 1995; Grupe et al.\ 1999).
%
% John -- I edited this sentence, but to be honest, it seems a little
% out of place to me.  Maybe things would flow better if we left it out?
%
% The constellation of characteristics associated with the NLS1 phenomenon
% may be linked to the mass and/or accretion rate of the black hole, the
% orientation of the nucleus to our line of sight, or a combination of
% these (e.g., Boller et al.\ 1996).
%
%\placefigure{figure3}

Remarkably, not only does a large fraction of the ROSAT-detected
Seyfert 1 sample of KISS galaxies exhibit narrow lines (7 out of 13,
or 54\%), but {\it a large fraction of all KISS Seyfert 1 galaxies 
are NLS1's}.  In total there are 16 Seyfert 1 galaxies 
in the KISS catalog within the H$\alpha$-selected redshift region
($z < 0.095$).  Of these, at least 7 (44\%) would qualify as NLS1's.
While the fraction of sources among the general population of Seyfert 1 
galaxies possessing narrow lines is not well known, the numbers for
the KISS sample (which is selected in a relatively unbiased way, at
least with regard to line widths) suggest that it may be quite high.
Further analysis of this issue will await more complete spectroscopic
follow-up of the KISS ELGs.

\subsection{Determination of $L_X$}
\label{refine}

In order to find the X-ray luminosities of the objects in the ROSAT-KISS sample, we first must convert RASS count rates to fluxes in the 0.1-2.4 keV band. To do this, we apply an absorbed power-law model with $\Gamma = 2.5, N_H =$ Galactic + $2.0\times 10^{20}$ $\mathrm{cm}^{-2}$ to convert PSPC source counts to X-ray flux. These spectral parameters are consistent with the methods of \citet{h01}, who determined $\Gamma = 2.5, N_H =$ Galactic + $2.0\times 10^{20}$ $\mathrm{cm}^{-2}$ through modeling of ROSAT PSPC spectra for a low-luminosity, distance-limited AGN sample. \citet{v99} use $\Gamma=2.3$, with $N_H$ equal to the Galactic value, for flux calculations (found in the electronic form of the RASS correlation catalogs). We make use of PIMMS\footnote{http://cxc.harvard.edu/toolkit/pimms.jsp; PIMMS software was originally developed at NASA/GSFC} in performing the conversion between ROSAT PSPC source counts and the $0.1-2.4$ keV flux. Combining the calculated X-ray fluxes from the above spectral model with the distances calculated from measured redshifts allows us to calculate the absorption-corrected $0.1-2.4$ keV luminosity of each source. This information is given in Table \ref{table2}. We use the \emph{colden}\footnote{http://cxc.harvard.edu/toolkit/colden.jsp} software package, in which column densities are drawn from the NRAO dataset \citep{dl90}, to calculate the Galactic H I column density along the line of sight to each galaxy. For the ROSAT-KISS objects in the $30^{\circ}$ strip, the Galactic component of $N_H$ has a median value of $1.57 \times 10^{20}$ $\mathrm{cm}^{-2}$; $43^{\circ}$ strip objects have a median Galactic component of $1.50 \times 10^{20}$ $\mathrm{cm}^{-2}$. 

\subsubsection{$L_X-L_{H\alpha}$ Relation}
\label{LxLHa}

%\placefigure{figure4}
\begin{figure*}[htp]
\vskip 0.2in
\epsfxsize=4.4in
\hskip 1.3in
\epsffile{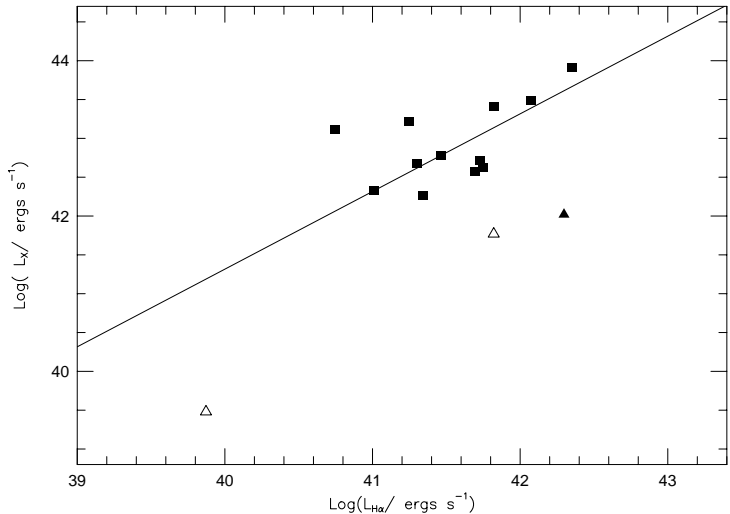}
\figcaption[/d7/kiss/rosat/paper/figures/xhaplot.eps]{The $L_X-L_{H\alpha}$ relation for the extragalactic ROSAT-KISS objects. Seyfert 1 galaxies are represented by filled squares, Seyfert 2 galaxies are shown as filled triangles, and the open triangles correspond the LINERs. The solid line corresponds to the median $L_X/L_{H\alpha}$ ratio of 20.7. \label{figure4}}
\end{figure*}

Knowledge of the total X-ray emissivity over the KISS volume is necessary for an understanding of the XRB contribution made by the low-luminosity AGNs in the KISS sample (Section 4). However, the majority of the KISS sample is not detected by ROSAT. Since the optical luminosities of the ROSAT-KISS sample are not drastically different from those of the general sample (Section \ref{Properties}), we expect that the contribution to the XRB made by KISS objects below the ROSAT threshold will be non-negligible. This implies that an estimate of the X-ray luminosity of the non-detected KISS galaxies is necessary for an accurate estimate of the XRB contribution from the full KISS AGN sample.

In an attempt to constrain the X-ray luminosities of the KISS Seyfert galaxies not included in the ROSAT-KISS sample, we derive an $L_X-L_{H\alpha}$ relation using the 15 lower redshift extragalactic objects for which X-ray data are available. The H$\alpha$ luminosities of these galaxies are calculated using the objective-prism spectrum of each galaxy.  The objective-prism fluxes are used, rather than the values from the follow-up slit spectra, because the former are calibrated in a uniform, homogeneous way (see KR1 and Gronwall et al. 2002c).  Since the $H\alpha$ and [N II] ($\lambda\lambda 6583, $6548 \AA) lines are blended in the objective-prism spectra, the observed [N II]/$H\alpha$ ratio from the follow-up spectra are used to correct the objective-prism fluxes for the contribution due to [N II].  In addition, the fluxes of the Seyfert 2 and LINER galaxies are corrected for intrinsic absorption using reddening coefficients calculated from the Balmer decrement. This latter correction can be especially significant for the LINERs in our sample. The resulting $L_X-L_{H\alpha}$ relation is shown in Figure 4.  In the figure Seyfert 1 galaxies are plotted as filled squares, the lone Seyfert 2 as a filled triangle,
and the two LINERs as open triangles.

There appears to be a fairly well-defined linear relation between $L_X$ and $L_{H\alpha}$ for these galaxies, at least for the Seyfert 1s. The median $L_X/L_{H\alpha}$ ratio for the volume-limited sample is 20.7, which is illustrated by the solid line in Figure 4.  Although a significant portion of the KISS AGNs have a soft $L_X$ too faint to allow detection by ROSAT, this does not eliminate the possibility that these objects may be responsible for a non-trivial XRB contribution.   

\subsubsection{Prediction of $L_X$ for objects with no X-ray detections}
\label{Predictions}

%\placefigure{figure5}
\begin{figure*}[htp]
\vskip 0.2in
\epsfxsize=4.4in
\hskip 1.3in
\epsffile{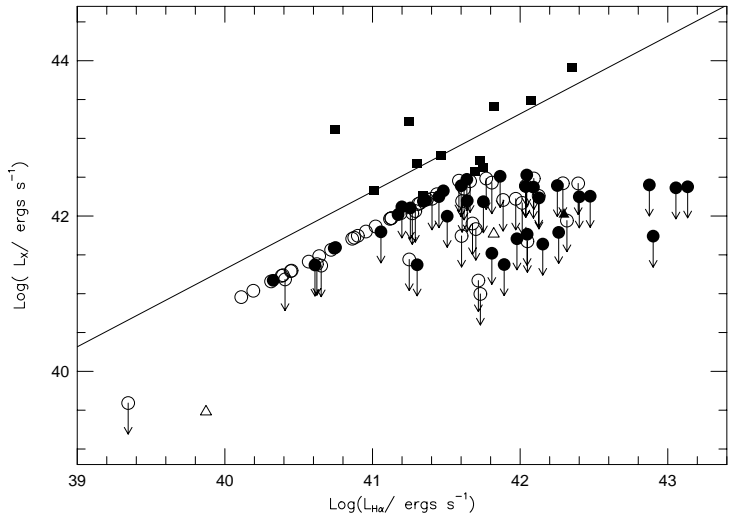}
\figcaption[/d7/kiss/rosat/paper/figures/finalplot.eps]{The $L_X/L_{H\alpha}$ plot for all KISS AGNs, with objects not detected by ROSAT having their X-ray luminosities estimated as described in the text.  Objects belonging to the ROSAT-KISS sample are represented as in Figure 4.  Non-detected Seyfert galaxies are plotted as filled circles, while non-detected LINERs are open circles. Upper-limit arrows are included for all objects whose $L_X$ estimate is based on the FSC flux limit (see text). \label{figure5}}
\end{figure*}

As mentioned above, many KISS AGNs have not been detected by RASS.  In order to derive an accurate assessment of the contribution of KISS-type AGNs to the soft XRB, the X-ray luminosities of the undetected objects must be estimated.  To do this, we employ the L$_X$/L$_{H\alpha}$ ratio and the measured H$\alpha$ luminosities of the galaxies. Our initial plan had been to utilize the value derived in the previous subsection for the ROSAT-KISS sample.  However, examination of Figure 4 shows that our median value for the $L_X/L_{H\alpha}$ ratio is dominated by the Seyfert 1 galaxies in the sample, while the Seyfert 2s and LINERs are systematically lower. Since the vast majority of non-ROSAT-detected objects are Seyfert 2s and LINERs, we wanted to use an $L_X/L_{H\alpha}$ ratio that was representative of these classes of AGN.  However, we did not feel comfortable using the small sample of narrow-lined ROSAT-KISS galaxies (three) to determine a median $L_X/L_{H\alpha}$ ratio to be used in our computations.  Therefore, we decided to adopt the $L_X/L_{H\alpha}$ ratio determined by \citet{h01} for their local sample of mainly narrow-lined AGNs.  These authors found a median $L_X/L_{H\alpha}$ of 7.

It is no great surprise that the median $L_X/L_{H\alpha}$ we calculate for the ROSAT-KISS sample is larger than that found by \citet{h01}, since our sample is dominated by Seyfert 1 galaxies.  Because of the greater obscuration expected for type 2 Seyferts in the dusty torus model, unabsorbed emitters such as Seyfert 1s are expected to have a significantly higher $L_X/L_{H\alpha}$ in the soft X-ray band.  This
is precisely what we observe in our sample.

One cannot simply use the value of the $L_X/L_{H\alpha}$ ratio to estimate $L_X$.  Examination of the distribution of FSC sources with flux reveals a loss of completeness beginning at approximately $0.015$ cts/s. We therefore adopt this value as an upper limit on the PSPC count rate for our non-detected objects. Computing $L_X$ using the average $L_X/L_{H\alpha}$ ratio described above will, in some cases, lead to an estimate of the X-ray flux for an object which lies above the FSC flux limit. We therefore adopt the upper limit $0.1-2.4$ keV count rate of 0.015 cts/s for all such objects, rather than the value inferred from the $L_X/L_{H\alpha}$ ratio. This method yields, in turn, an upper limit for the soft XRB contribution of these non-ROSAT-detected KISS galaxies. Figure \ref{figure5} shows the $L_X - L_{H\alpha}$ relation with the non-ROSAT KISS galaxies included.  The non-detected Seyfert galaxies are plotted as filled circles, while non-detected LINERs are open circles.  Objects whose $L_X$ estimate is based on the FSC flux limit, as descibed above, are indicated by an upper-limit arrow.  Many of the non-detected Seyfert galaxies fall into this group.  All other non-detections have had their X-ray luminosities predicted using the relation $L_X=7\cdot L_{H\alpha}$.  These objects, which include many of the LINERs, fall along a straight line in the plot.

There are a total of 65 Seyfert galaxies in the KISS database. Fourteen of these galaxies are detected by ROSAT, leaving 51 Seyferts with no X-ray flux data. After elimination of those galaxies outside the KISS completeness limit ($z=0.0875$; see Section 4.1), 36 non-ROSAT-detected Seyfert galaxies remain. KISS also detects a large population of LINERs - there are 55 LINERs in the combined $30^{\circ}$ and $43^{\circ}$ KISS database. Most of these objects have not been detected by ROSAT, and nearly all of them lie within the $z=0.0875$ completeness limit; we predict $L_X$ using the above method for 52 LINERs in total.

\subsection{Individual objects}
\label{indiv}
Since the ROSAT-KISS sample consists of just 18 objects, it is possible for us to examine each one in some detail, while still making generalizations about the sample as a whole. This section details the properties of selected ROSAT-KISS objects.

\emph{KISSR 29:} There are only two LINERs in the ROSAT-KISS sample (the other example being KISSR 2059). KISSR 29 (NGC 4278) is remarkable in that it is one of the optically brightest sources in the KISS survey ($m_B = 11.53$); this galaxy is extremely nearby (only 8 Mpc distant). KISSR 29 is the only ROSAT-KISS object for which archival HRI images are available. However, in the interests of uniformity we have not made use of the pointed data in our calculations. \citet{h01} use pointed PSPC observations of KISSR 29 to derive an estimate of the total soft X-ray flux (0.1-2.4 keV) of $5.92 \times 10^{-13}$ ergs $\mathrm{s}^{-1}$ $\mathrm{cm}^{-2}$, assuming a photon index of $\Gamma=2.5$ and column density $N_H=$ Galactic + $2.0 \times 10^{20}$ $\mathrm{cm}^{-2}$. Using the same spectral parameters, we find that the 0.1-2.4 keV flux for this object is $2.49 \times 10^{-13}$ ergs $\mathrm{s}^{-1}$ $\mathrm{cm}^{-2}$ using the RASS detection. 

\emph{KISSR 817:} This is the only stellar X-ray source in our sample. KISSR 817 is believed to be a cataclysmic variable star serendipitously detected by the All-Sky Survey; it was detected as part of our sample due to its relatively high soft X-ray flux. The follow-up optical spectrum detected no $H{\alpha}$ emission. Its spectrum resembles that of a hot white dwarf, and it is quite blue in color ($B-V=-0.09$). This is consistent with the behavior of a cataclysmic variable; the star may have been in a high state when the objective-prism survey data were gathered. This would explain the presence of the emission line in the survey data, of which no evidence was seen in follow-up observations.

\emph{KISSR 864:} This object is one of the five QSOs identified to date in the KISS catalog. As with all high-redshift KISS objects, it was detected via the presence of a strong emission line other than H$\alpha$ which had redshifted into the relevant wavelength range. In this case, the emission line was $H\gamma$. KISSR 864 is the most luminous object in the ROSAT-KISS sample, both in the optical ($M_B = -24.33$) and in the X-ray ($log(L_X)=44.61$).

\emph{KISSR 1494:} One KISS Seyfert 2 galaxy, KISSR 1494, was detected by the All-Sky Survey. This galaxy is one of the more luminous sources in our ROSAT sample, with an absolute magnitude of -20.89. The median absolute magnitude for the overall sample of 46 KISS Seyfert 2 galaxies is $M_B = -19.76$.

\emph{KISSR 1802}: This object is the only high-redshift Seyfert 1 galaxy detected by ROSAT-KISS. The H$\beta$ and [O III] emission lines are redshifted into the KISS bandpass, revealing a redshift of $z=0.4213$.  As one might expect, this object is a very powerful X-ray source, with $log(L_X)=44.20$, making it the second most luminous source in the ROSAT-KISS sample.

\emph{Seyfert 1 objects:} For our purposes, all galaxies exhibiting broad Balmer emission lines and with $M_B \geq -23$ are termed Seyfert 1 galaxies. This includes a few ``intermediate'' Seyferts, such as the Seyfert 1.5 galaxies KISSR 1503 and KISSR 1802. The thirteen Seyfert 1 galaxies in the ROSAT-KISS sample have a very high median luminosity: $M_B= -20.16$ as compared to the general median $M_B=-18.86$ for all KISS objects. The median $M_B$ for the ROSAT-KISS Sy 1's is statistically indistinguishable from the value for the general KISS Sy 1 sample. As discussed above, the proportion of narrow-line Seyfert galaxies (H$\beta$ FWHM $\le 1500$ km/s, [O III]/H$\beta \le$ 3) in the ROSAT-KISS sample seems to be significantly higher than that found in previous optically-selected AGN samples.

\section{Contribution to the X-ray Background}

Determination of the relative contributions of various source populations to the X-ray background is of major importance.  As summarized in Section 1, a modest fraction of the soft XRB may well be due to a population of moderate-luminosity AGNs.  In this section, we perform comoving volume and emissivity calculations in order to estimate the XRB contribution from the types of AGNs detected locally by KISS. We assume a cosmology of $H_0=75$ $\mathrm{km/s/Mpc}$, $\Omega_M=1.0$, $\Omega_\Lambda=0.0$ for all computations.

\subsection{Comoving volume determination}
\label{vol}

Computation of the total volume covered by KISS (necessary for volume emissivity determination; see Section \ref{XRB}) requires us to set a maximum redshift for the survey. The redshift cutoff for KISS objects does not occur instantaneously: we experience significant ``edge effects'' for objects near the cutoff redshift, due to loss of flux from the relevant emission line (H$\alpha$). Studies \citep{gron02c} indicate that this effect begins to become significant at approximately $z = 0.0875$. We therefore do not consider any KISS objects with spectroscopic redshifts larger than $z=0.0875$. This results in elimination of one source (KISSR 1903; z=0.094). In addition, we do not consider the two high-redshift objects noted in Section \ref{indiv} (KISSR 864 and 1802) during computation of volume emissivities.

Edge effects are present not only at the outskirts of a survey volume, but at the inner limit as well; the distribution of extremely nearby sources will be heavily influenced by the local structure of the Universe. We have therefore excluded from our calculations the only ROSAT-KISS source close enough to experience these effects - the LINER, KISSR 29 (NGC 4278). 

After the elimination of all sources outside the relevant volume, we are left with a sample of 13 galaxies. This sample, selected from a nearly volume-limited survey, should be representative of the local X-ray - emitting galaxy population as a whole.

\subsection{Volume emissivity and XRB calculations}
\label{XRB}

Accurate estimates of the emissivity produced by the objects in the KISS volumes are necessary to obtain a knowledge of the XRB contribution from these objects. Once the emissivity over the actual volume is known, it can be generalized to all redshifts using the appropriate evolution model (see below) in order to estimate the X-ray intensity over the sky area in question, which can then be compared with the known intensity of the XRB. The areal coverage of the KISS strips is, in fact, fairly small: the $30^{\circ}$ strip covers an area of 62.16 $\mathrm{deg}^{2}$, while the $43^{\circ}$ strip covers 65.86 $\mathrm{deg}^{2}$. The volume encompassed by the survey is nevertheless quite large relative to magnitude-limited surveys, because of the comparatively large redshift limit. The comoving distance limit corresponding to our adopted $z_{max} = 0.0875$ is $\approx$ 343 Mpc. This leads to volume emissivities in the $0.1-2.4$ keV band of $\rho = (6.63 \pm 2.34) \times 10^{38}$ ergs $\mathrm{s}^{-1}$ $\mathrm{Mpc}^{-3}$ for the $30^{\circ}$ strip and  $\rho = (1.45 \pm 0.65) \times 10^{38}$ ergs $\mathrm{s}^{-1}$ $\mathrm{Mpc}^{-3}$ for the $43^{\circ}$ strip.  The quoted errors are simply Poissonian, based on the numbers of objects detected in the two survey strips.  We have opted to employ Poisson statistics in computing our uncertainties since small number statistics will dominate over the uncertainties in deriving the X-ray luminosities for the individual objects.  Our approach leads to more conservative estimates of the uncertainties.  It is worthwhile to note that the volume emissivities from the two KISS strips differ by a factor of over four. This is largely a result of the small number of objects in each sample. In fact, much of the difference between the XRB contributions from the ROSAT-KISS objects in the two strips is due to a single, high-luminosity source in the $30^{\circ}$ strip (KISSR 924). 

The above numbers represent the luminosity contributions from the ROSAT-detected subsample of the KISS objects. However, as discussed in Section \ref{Predictions}, it is possible that those objects which are too faint to have been detected by ROSAT may nonetheless be significant contributors to the total volume emissivity. Using X-ray luminosities estimated from the observed H$\alpha$ luminosities and the adopted $L_X/L_{H\alpha}$ ratio, we find that inclusion of the non-detected KISS galaxies raises the volume emissivity to $(10.24 \pm 1.26) \times 10^{38}$ ergs $\mathrm{s}^{-1}$ $\mathrm{Mpc}^{-3}$ in the $30^{\circ}$ strip, and to $(2.76 \pm 0.46) \times 10^{38}$ ergs $\mathrm{s}^{-1}$ $\mathrm{Mpc}^{-3}$ in the $43^{\circ}$ strip.  Again, the errors are Poissonian in nature, based on N = 66 AGNs in the $30^{\circ}$ strip and N = 36 in the $43^{\circ}$ strip.  The inclusion of objects below the ROSAT detection threshold increases the volume emissivity over the KISS volumes by nearly a factor of 2, indicating that these objects do indeed have a significant effect on the global soft X-ray emission from KISS-type objects. Of course, because of the prediction method employed, the above values represent upper limits on the volume emissivity from the KISS galaxies spectroscopically identified as AGNs.

One additional concern remains to be considered. Since we do not have follow-up optical spectra for all galaxies in the KISS database, selection effects are introduced into our XRB calculations. Currently, 64\% and 14\% of the ELG sample possess follow-up spectra in the $30^{\circ}$ and $43^{\circ}$ strips, respectively. This incompleteness results in an underestimate of the XRB contribution from the KISS ELGs. We use the properties of the $30^{\circ}$ and $43^{\circ}$ samples with follow-up spectra to correct for incompleteness effects. Assuming the proportion of AGNs in the total galaxian population is constant, we use the observed number of galaxies with follow-up spectra to correct for the number of additional AGNs expected after completion of spectral identification. After performing this correction, we find that the total volume emissivity contributions predicted from the inclusion of non-RASS detected Seyfert galaxies and LINERs in the $30^{\circ}$ and $43^{\circ}$ strips are $(11.76 \pm 1.45) \times 10^{38}$ ergs $\mathrm{s}^{-1}$ $\mathrm{Mpc}^{-3}$ and $(5.01 \pm 0.84) \times 10^{38}$ ergs $\mathrm{s}^{-1}$ $\mathrm{Mpc}^{-3}$, respectively.  The errors quoted are based on the Poissonian uncertainties for the known AGNs in the two strips, as indicated in the preceding paragraph.

Given these results for the emissivity within the survey volumes, we then calculate the X-ray intensities using the results of \citet{s96}:

\[I_X=\frac{c\rho}{4 \pi H_0}\int\limits_{0}^{\infty}\frac{(1+z)^{2-\Gamma}}{(1+z)^{3} \sqrt{1+\Omega_M z} }\mathrm{d}z\]
We have ignored the possibility of redshift evolution of the X-ray luminosity function (XLF) in our calculations, which introduces significant uncertainties into our result. \citet{gilli00} and \citet{m00, m01} use population synthesis models to calculate the evolution of the soft X-ray AGN luminosity function from ROSAT data. The inclusion of evolutionary effects will increase the integrated X-ray intensity (see below). The fact that upper limits have been used to predict the total X-ray intensity over the KISS areas due to objects not detected by ROSAT will tend to compensate for the lack of consideration of evolution in our calculations, but it is not possible to determine precisely to what degree this is true.

 Using the power-law model with spectral parameters discussed in Section \ref{refine}, we then convert the 0.1-2.4 keV intensities from each survey strip to the 0.5-2 keV band. We can then apply the result of \citet{s96}, namely, that the total 0.5-2 keV XRB intensity is $2.61 \times 10^{-8}$ ergs $\mathrm{s}^{-1}$ $\mathrm{cm}^{-2}$ $\mathrm{sr}^{-1}$. Using this value for the XRB intensity along with our corrected volume emissivities, we find that the total XRB contribution in the $30^{\circ}$ strip is $31.1 \pm 3.8$\%, and is $13.2 \pm 2.2$\% in the $43^{\circ}$ strip. The average of these two values is $\mathbf{22.1 \pm 8.9\%}$.
The large uncertainty we quote in this final estimate reflects the scatter between the measurements from the two survey areas, rather than the Poissonian uncertainty used in the errors for the XRB contributions from the individual survey strips.  We feel that this larger value is likely to be more indicative of the actual uncertainty in our result.  We stress that the predicted XRB contribution for the types of AGNs detected in local samples like KISS does not take into account evolution of the AGN population with lookback time.  However, \citet{h01} found that application of the pure density evolution model derived for luminous Seyfert 1s by \citet{m00} leads to an {\it overproduction} of the 0.5-2 keV XRB by low-luminosity AGNs, given the non-evolving contribution of 9-11\% they measured for such objects. Since the XRB contribution found for the ROSAT-KISS sample is somewhat higher than the Halderson et al. value, application of the Miyaji evolution model will also overproduce the XRB for our sample in this energy range. While the result from this particular evolutionary model is not useful for tightening estimates of the XRB produced by low-luminosity AGNs, it should serve as a mild constraint on models of the evolution of X-ray luminosity from KISS-type objects. The Miyaji et al. models are based primarily on high luminosity sources, for which the evolution with redshift may well differ from that of galaxies with lower $L_X$.

\section{Discussion}

The depth and completeness of KISS allows the survey to make unique contributions to the study of emission-line galaxies. Due to its nearly volume-limited nature, this sample will allow for an accurate determination of the statistical properties of AGNs. The global properties of the KISS sample should generalize with a reasonable degree of accuracy to the low-redshift AGN population as a whole. Thus we can use the (statistically small) sample of ROSAT-detected KISS AGNs to study the general properties of local active galaxies. 

The use of the ROSAT All-Sky Survey in correlation with KISS gives rise to certain selection effects. ROSAT focuses on the soft ($0.1-2.4$ keV) X-ray band.  Thus, it is to be expected that such a survey will preferentially detect larger numbers of unabsorbed Seyfert 1 galaxies. This expectation is reflected in the ROSAT-KISS sample. Nearly all of the 19 Seyfert 1 galaxies present in the KISS sample have been detected by the All-Sky Survey; only 6 KISS Sy 1's are not. Only a very small fraction of the known KISS narrow-line AGNs (Seyfert 2s and LINERs) are ROSAT-detected. In addition, none of the KISS starburst galaxies are detected.

Our estimate for the soft XRB contribution by ROSAT-KISS objects is roughly a factor of two higher than that found by \citet{h01} for lower-luminosity AGNs with $L_X = 10^{38} - 10^{40}$ ergs $\mathrm{s}^{-1}$.  The different values for the soft-XRB contributions found for the two studies are due, in part, to the fact that the KISS AGNs are on average more luminous (both optically and in the X-rays) than the galaxies in the Halderson et al. sample.  The KISS AGNs are detected over a much larger volume than the very local Halderson et al. sample, meaning that the intrinsically rare luminous AGNs have a much better chance of showing up in KISS.  Another factor is that, unlike Halderson et al. who worked with a fairly complete sample of X-ray-detected AGNs, our calculations of the total volume emissivity and soft XRB contribution from KISS galaxies include predictions (often upper limits) of the X-ray luminosities of non-detected KISS objects.  The inclusion of non-detected Seyfert and LINER-type objects in calculations of the XRB contribution from ROSAT-KISS leads to a significant increase in our derived value. The total contribution to the 0.5-2.0 keV XRB is approximately $22.1 \pm 8.9\%$ after corrections have been performed, representing the average of the values from the $30^{\circ}$ and $43^{\circ}$ strips ($31.1 \pm 3.8\%$ and $13.2 \pm 2.2\%$, respectively). The difference between the XRB contributions from the $30^{\circ}$ and $43^{\circ}$ strips remains large (more than a factor of 2) even after correcting for the incomplete nature of KISS follow-up spectral data. This variation may be due in part to differences in large-scale structure between these two regions of the sky, but is primarily due to the presence of one or two very luminous sources in the $30^{\circ}$ strip, with no corresponding luminous objects detected in the $43^{\circ}$ strip. 

Our average XRB contribution represents an upper limit to the 0.5-2 keV X-ray XRB contribution since a significant number of the KISS objects were not detected by ROSAT and were therefore assigned upper-limit values for $L_X$. Our results are consistent with recent research showing that high-redshift, more luminous AGNs are the dominant source of XRB emission in both the hard and soft energy bands. Although objects like those in the KISS catalog cannot be the dominant source of the soft XRB, it is possible that their contribution may become much more important when viewed in hard X-rays, where the effects of absorption are far less significant.

Because KISS is a local sample of galaxies, it is expected that the AGNs in KISS will be less luminous than a similar sample at higher redshift, due to a combination of sensitivity and evolution effects. We have not included such evolution in our calculations, a fact which may have significant impact on our results. Several authors \citep{gilli00, m00, m01, gr96} show that the luminosity function of AGNs does in fact evolve significantly with redshift, and \citet{m01} succeed in correcting for evolutionary effects when generalizing local volume emissivity to XRB contribution. However, there is some evidence that the evolution with redshift is strongest for the most luminous sources, and hence may not be directly applicable to our sample. In the absence of an evolutionary model which does not overproduce the XRB when applied to the ROSAT-KISS sample, our estimate for the soft XRB contribution is a reasonable first-order result.

\section{Conclusions}

We have conducted a positional correlation between the KPNO International Spectroscopic Survey and the ROSAT All-Sky Survey databases. There are 18 non-spurious matches in total, 72\% (13) of which are spectroscopically identified as Seyfert 1 objects using follow-up spectral observations. Accurate, high-resolution optical spectra are available for all ROSAT-KISS objects.

The fact that most soft X-ray detected KISS objects are identified as Seyfert 1 galaxies is explained \emph{via} selection effects. Type 1 Seyferts, being less highly absorbed, are expected to be far more luminous in the soft X-ray, and their preferential detection by ROSAT-KISS comes as a natural consequence of the sample selection. A large proportion (54\%) of the Seyfert 1s in the ROSAT-KISS sample are in fact narrow-line Seyfert 1 galaxies; this is a reflection of the high fraction of NLS1s in the general KISS sample (44\% of Seyfert 1s with follow-up spectra available to date).
 
Notable non-Seyfert 1 objects within the ROSAT-KISS sample include one Seyfert 2 galaxy and two LINERs; the LINERs are intrinsically faint objects, which  may have been detected due to their proximity. Several ROSAT-KISS sources lie outside our adopted volume limits; these objects have been discarded from subsequent calculations. ROSAT-KISS includes one of the five QSOs in the KISS database, KISSR 864. We also note the presence of a stellar object, KISSR 817, which is bright in the soft X-ray.

KISS provides a nearly complete sample of ELGs over a well-defined volume in space. We therefore expect that the XRB contributions estimated from the ROSAT-KISS correlation will yield an accurate representation of the local contribution of AGNs to the soft XRB. The locality of the KISS volume may imply inaccuracies in our result due to the fact that evolution has not been taken into account, but application of the density evolution model of \citet{m00} result in overproduction of the soft XRB. We estimate upper limits for the 0.1-2.4 keV $L_X$ for KISS objects not detected as X-ray sources by ROSAT, using the median $L_X/L_{H\alpha}$ from a previous study. We also predict the number of additional AGNs expected to be identified after follow-up spectra have been obtained for all KISS objects, and adjust the emissivity accordingly. We find that the upper limit for the XRB contribution from KISS AGNs is $22.1 \pm 8.9\%$, the average over the two KISS areas. This result is a factor of two higher than the 9-11\% contribution estimated by \citet{h01} for low-luminosity AGNs.  The differences between the two studies is primarily the depth, and hence luminosity ranges, sampled by the two surveys.  It should be remembered that the KISS-dervied XRB contribution is an upper limit, due to the manner in which the X-ray luminosities are estimated for our non-ROSAT-detected sources.

This project has taken a different approach to the study of the XRB. Instead of studying the optical properties of an X-ray selected sample, we examine the X-ray properties of a sample of optically luminous galaxies. Future studies of the global X-ray emission properties of volume-limited galaxy samples may prove highly significant in constraining the XRB contribution from individual galaxian populations, as well as the redshift evolution of the X-ray luminosity function. The KISS AGN sample would, in fact, be an excellent one to use for establishing the X-ray characteristics for the low-redshift, low- and intermediate-luminosity AGN population. Our sample size is large enough to construct an X-ray luminosity function to be used as a template for future studies of the evolution of X-ray sources. Currently, the biggest limitation is the lack of quality X-ray observations for the bulk of the KISS AGN sample.

\acknowledgments

We gratefully acknowledge financial support for the KISS project from an NSF 
Presidential Faculty Award to JJS (NSF-AST-9553020), as well as continued 
support for our ongoing follow-up spectroscopy campaign (NSF-AST-0071114).  
We also thank Wesleyan University for providing additional funding for several 
of the observing runs during which the important follow-up spectral data were 
obtained.  We thank the numerous KISS team members who have participated in 
these spectroscopic follow-up observations during the past few years, 
particularly Caryl Gronwall, Drew Phillips, Gary Wegner, Laura Chomiuk, Jason 
Melbourne, Jeff Van Duyne, Kerrie McKinstry, and Robin Ciardullo.  We appreciate 
help from Eli Beckerman and Paul Green regarding accessing the ROSAT data as well 
as for discussions concerning the x-ray properties of AGN.  Several useful 
suggestions by the anonymous referee helped to improve the presentation of this paper.  
Finally, we wish to thank the support staffs of Kitt Peak National Observatory,
Lick Observatory, the Hobbey-Eberly Telescope, MDM Observatory, and Apache
Point Observatory for their excellent assistance in obtaining both the survey
data as well as the spectroscopic observations.  Without their assistance
this project would have been impossible.

\vskip -1.0in
\renewcommand{\arraystretch}{.6}
\begin{deluxetable}{crrrrrrrrrrr}
\footnotesize
\tablecaption{The ROSAT-KISS matching. \label{table1}}
\tablewidth{0pt}
\tablehead{
\colhead{} & \colhead{BSC}   & \colhead{FSC} & \colhead{Total}
}
\startdata
KISS $30^{\circ}$ strip (KR1) &5 &6 &11 \\
KISS $43^{\circ}$ strip (KR2) &2 &5 &7 \\
\tableline
Total &7 &11 &18 \\
 \enddata

\end{deluxetable}

\clearpage

\renewcommand{\arraystretch}{.6}
\begin{deluxetable}{rrllrcrrrrrr}
\tablecolumns{12}
\scriptsize
\rotate
\tablecaption{The ROSAT-KISS sample. \label{table2}}
\tablewidth{0pt}
\tablehead{
\colhead{KISSR} & \colhead{RASS name}   & \colhead{RA} & \colhead{Dec} & \colhead{$\Delta$} & \colhead{ELG} & \colhead{$M_B$} & \colhead{$m_B$} & \colhead{B - V} & \colhead{Redshift} & \colhead{$f_X$} & \colhead{$log(L_X)$} 
\\
\colhead{\#} && \colhead{(J2000)} & \colhead{(J2000)} & \colhead{''} & \colhead{type} &&&&& \colhead{cts/s} & \colhead{ergs/s}
\\ 
\colhead{(1)} & \colhead{(2)} &  \colhead{(3)} &\colhead{(4)} & \colhead{(5)} & \colhead{(6)} & \colhead{(7)} & \colhead{(8)} & \colhead{(9)} & \colhead{(10)} & \colhead{(11)} & \colhead{(12)} 
}
\startdata
29&J122006.2+291643&12:20:06.8&29:16:50.3&12.1&LINER&-18.06&11.53&0.99&0.00203&0.036&39.48\\
281&J132438.5+291020&13:24:38.7&29:10:11.9&8.6&Sy 1&-19.43&17.95&0.72&0.07257&0.050&42.72\\
350&J134607.5+293814&13:46:08.1&29:38:10.7&9.5&Sy 1&-20.97&16.51&0.47&0.07663&0.261&43.49\\
360&J135020.1+291005&13:50:19.7&29:10:03.2&6.3&Sy 1&-18.05&19.13&1.20&0.06680&0.021&42.27\\
489&J141434.6+293439&14:14:34.5&29:34:28.7&10.4&Sy 1&-19.61&17.85&0.50&0.07590&0.041&42.68\\
531&J142219.8+294257&14:22:20.2&29:42:55.5&6.1&Sy 1&-20.67&15.99&0.60&0.05289&0.105&42.78\\
583&J145109.1+292633&14:51:09.4&29:26:26.0&8.0&Sy 1&-20.19&17.30&1.22&0.07563&0.137&43.22\\
632&J150620.7+294005&15:06:20.8&29:39:57.3&8.4&Sy 1&-19.57&17.36&0.93&0.05895&0.028&42.33\\
817&J153619.3+292146&15:36:18.6&29:21:50.9&11.4&CV (?)&.......&19.34&-0.09&0.00000&0.058&.......\\
864&J154842.8+290901&15:48:43.7&29:08:59.8&13.3&QSO&-24.33&17.70&0.06&0.55010&0.049&44.61\\
924&J155643.0+294838&15:56:42.9&29:48:48.0&9.7&Sy 1&-21.76&16.08&0.68&0.08554&0.457&43.91\\
1494&J131325.3+433219&13:13:25.8&43:32:14.2&9.6&Sy 2&-20.89&15.95&0.82&0.05732&0.015&42.02\\
1503&J131406.3+434649&13:14:07.1&43:46:34.1&19.7&Sy 1&-19.40&18.36&1.14&0.08723&0.026&42.62\\
1511&J131511.5+432544&13:15:10.1&43:25:46.8&21.8&Sy 1&-20.16&17.58&0.62&0.08614&0.024&42.58\\
1755&J141755.1+431152&14:17:55.5&43:11:55.2&6.8&Sy 1&-19.90&17.66&0.57&0.07902&0.201&43.41\\
1802&J143239.7+430021&14:32:41.1&43:00:39.8&28.5&Sy 1&-21.78&19.57&0.11&0.42133&0.038&44.20\\
1903&J150506.8+435002&15:05:07.3&43:50:05.2&8.3&Sy 1&-20.21&17.74&1.00&0.09359&0.069&43.12\\
2059&J155339.7+434410&15:53:40.4&43:44:04.3&11.9&LINER&-19.67&16.48&0.90&0.04048&0.017&41.77\\
 \enddata

\end{deluxetable}

\end{document}